# Continuous and Pulsed Experiments with Numerical Simulation to Dissect Pituitary Gland Tumour by Using Liquid Jet

Ahmed H. Alamoud, Emmanuel Baillot, Camila Belabbas, Hadi S. Samimi Ardestani, Hamid Bahai, Adrien Baldit, Mahmoud Chizari

*Abstract*— Endoscopic endonasal surgery is a minimal invasive surgery that has been used to dissect pituitary gland tumour via curettes with the help of endoscope. However, this type of surgery has a high risk of failure because curettes may cause damages to blood vessels and optical nerves that lead to more complication for the patient. The aim of this study is to develop a new technique to dissect the tumour by using liquid jet. A series of experimental tests have been performed on animal tissue to study the effect of liquid pressure and nozzle diameter on dissecting and cutting the tissue. Continuous / pulsed liquid jet used with variable nozzle diameters, distances, pressures and angles. The study concluded with promising results on liquid jet to dissect hypophysis tumour and preserving fine blood vessels and optical nerve located near the pituitary gland.

## I. INTRODUCTION

One part of the body that is affected by both type of tumour is the brain, which is a crucial region of the body that controls all its functions. According to the latest statistics collected by Cancer Research UK, there were 9365 new cases of brain cancer reported in 2011, where more than 50% of the new cases number dies every year [1]. From these numbers, around 8% of brain tumours are in the pituitary gland, which is a serious medical condition lead to death if not treated [2], [3]. In the United States, the number of patients diagnosed with pituitary gland tumour is around 10,000 patients each year, in which most of these lesions are benign [4].

The skull base is divided into anterior, middle, and posterior regions that sited behind the eyes and nose, in which the tumours are categorized according to the involved area [5]. Therefore, treating skull base tumours are difficult since they are very near to crucial blood vessels and nerves that can cause health issues [6]. In addition, when trying to remove it surgically, it may cause, for instance: infection, hematoma, cerebrospinal fluid leak, and the most significant problem is difficulty to reach the lesion in a safe, quick and complete access [5]-[7]. One of the vital tumour in this region and can cause a severe health problem to the patient when not removed properly is hypophysis tumour, which is located in the skull base and affects the pituitary gland, also called pituitary gland tumour, which is responsible for secretion the main body hormones [8]. Thus, the tumour disorders the normal balance of hormones in the body that cause the patient to become sick [8]. Moreover, malignant type of this tumour affected only few patients [8]. Nevertheless, due to its location, the tumour develops upward, and consequently hypophysis lesion will cause vision problems when it reached and pressed on the optic nerve [8].

Surgery is considered the main treatment for hypophysis tumour and it depends on size, exact location and type of the tumour and whether the tumour is spread to neighbour tissues [9]. Since the pituitary gland is located at the back wall of the sinus, the common technique selected for more than 90% of patients to remove pituitary tumour is transsphenoidal surgery, which is performed from the nose via opening the sphenoid sinus bony walls with small surgical instruments [9]. Moreover, the current minimal invasive technique is done by using endoscope, which is a small camera attached at the tip of lighted thin fibre optic tube that allows the surgeon to do the operation with a small incision to reach the gland and remove the tumour [9].

In recent years a few attempt has been made to remove tumour using liquid jet. Ogawa et al. recently studied on a new technique to remove the hypophysis tumour [10]. This technique was applied by pulsed laser-induced liquid jet (LILJ) system that removes the tumour efficiently and safely without any harm to blood vessels and nerves. Moreover, a study has been published in 2015 by Nakagawa et al. regarding the safety of using LILJ system, which concluded that water jet is safe in removing lesions on pituitary gland and surrounding area [11]. As a result, more studies are needed to be conducted regarding water jet technology with different nozzle diameters and methods to create the water

Manuscript received May 08, 2017.

Ahmed H. Alamoud (corresponding author) is with Mechanical, Aerospace and Civil Engineering Department, College of Engineering, Design and Physical Sciences, Brunel University London, Uxbridge, UB8 3PH, UK, Phone: +447742167248; E-mail: ahmedhalamoud@gmail.com

Emmanuel Baillot is with National Engineering School of Metz (ENIM), Metz, France

Camila Belabbas is with National Engineering School of Metz (ENIM), Metz, France

Hadi S. Samimi Ardestani is with Tehran University of Medical Science, Research centre of otolaryngology, head and neck surgery, Amiralam Hospital, Tehran

Hamid Bahai is with Mechanical, Aerospace and Civil Engineering Department, College of Engineering, Design and Physical Sciences, Brunel University London, UK

Adrien Baldit is with National Engineering School of Metz (ENIM), Department of Mechanics, Physics, Mathematics, 1 route d'Ars Laquenexy, 57078 Metz Cedex 3, France

Mahmoud Chizari is with the School of Mechanical Engineering, Sharif University of Technology in Tehran. He is also with Mechanical, Aerospace and Civil Engineering Department, College of Engineering, Design and Physical Sciences, Brunel University London, UK jet stream and its effect on removing pituitary tumour in a minimal invasive procedure.

*A. Objective of the study*

The study aims to study the use of liquid jet on animal soft tissue. A series of experiments have been performed on continuous and pulsed liquid jet with various nozzle diameters to study its effect on cutting or peeling of the soft tissue. Along with the experimental study a numerical simulation has been done to evaluate the effect of liquid jet with variable nozzle diameters on cutting soft tissue.

## II. METHODOLOGY

The study is based on quantitative data taken from two set of experiments and a computer simulation. The first experiment is related to the effect of continuous waterjet with various nozzles diameter on cutting a chicken breast and the second experiment is to study a pulsed water jet on peeling the chicken breast soft tissue. Then, a simple simulation is conducted using Abaqus 6.14, which is finite element analysis (FEA) software, to study the effect of waterjet on penetrating a chicken breast soft tissue.

*A. Continuous waterjet with various diameters*

To measure the effect of continuous jet with various diameters on soft tissue; the experiment was conducted simply by using a water tab and designing a nozzle that fitted to the outlet of the tab. A manometer has been used to measure the water pressure that had a value of 3.2 bar ± 0.2.

Thus, 20 pieces have been created and a nozzle hole has been drilled for each piece via a variable speed rotary tool with drill bits that has a diameter range from 0.3 mm to 1.6 mm, which resulted in 14 circles.

After that, the soft tissue has been placed on a plexiglass board and fixed with rubber bands to avoid any movement caused by the flow of water. Furthermore, the distant between the nozzle and the soft tissue is 66 mm, which is constant for all samples (Fig. 1).

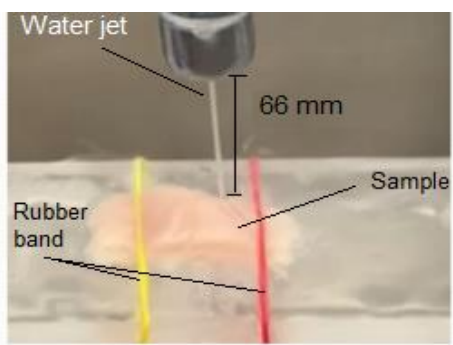

Fig. 1 Setup of the first experiment

*B. Pulsed liquid jet experiment*

The second experiment is conducted by a pulsed liquid jet to peel a sample of soft tissue from a chicken breast that has been done with 4 variables, which are angle and distant between the nozzle and the sample, liquid jet pressure, and nozzle diameter, to study the effect of these parameter on peeling soft tissue.

To do this test, the following parts were needed: First, a water pump with a reservoir to create a variable nozzle pressure. Second, a flexible hose that used to freely adjust the angle and a changeable nozzle to have more than one diameter. Therefore, to avoid building a new device from scratch and save time, a search has been done to find a suitable device that meet the requirements with less modification, which resulted in buying a Waterpik® Ultra Water Flosser that has 10 adjustable pressure settings with 1200 pulses per minute and a changeable nozzle [12]. The nozzle has been created by using a 3D fused deposition modelling (FDM) printer, Dimension Elite 3D from Stratasys Ltd., with production-grade thermoplastic ABSplus-P430 that provides a precise measurement with a layer thickness of 0.178 mm to make multiple nozzles with various diameters in a short time [13].

Then, the soft tissue has been placed on a plexiglass board that can be adjusted forward and backward according to the distance setting. The soft tissue then fixed with rubber bands on the plexiglass board. In addition, the distance between the nozzle and the sample was adjusted by the nozzle holder. The experimental setup is shown in Fig 2.

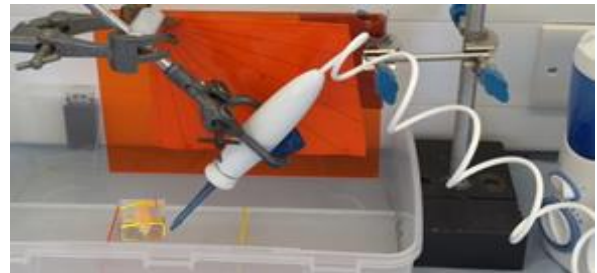

Fig 2 Setup of the second experiment

*C. Computer simulations*

To study the effect of waterjet on penetrating chicken breast soft tissue, a 2D model was created using finite element analysis software, Abaqus/CAE 6.14, which is capable of analysing the model and evaluate results [14].

The model carefully created in the software to get accurate results. To create the model firstly parts created. Then defining a property for each part, assemble parts, create step and a field output, define the interaction between parts, choose the load and boundary condition, mesh each part, create a job and submit it, monitor the job for any error, have respectively done to get the results. The model, as shown in Fig. 3, has been created for each waterjet diameter; 0.3 mm, 0.4 mm, 0.5 mm, 0.6 mm and 0.7 mm, which each has been simulated separately.

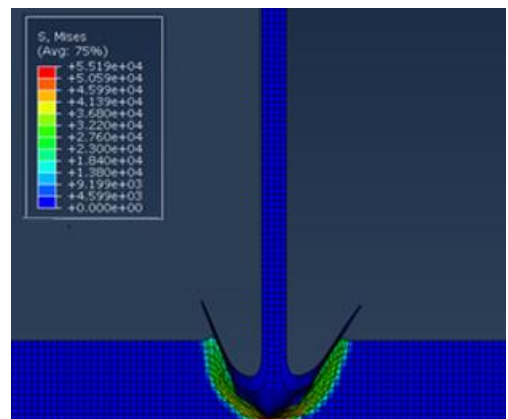

Fig. 3 A two-dimensional FEA model of the liquid jet penetrating into the soft target

In two-dimensional FEA modelling, a 2D deformable shell part representing the water jet has been sketched. The nozzle diameter was related to the waterjet width and the height of waterjet column was set to 30 mm. Also, another 2D deformable shell part representing the chicken soft tissue with geometry of 1mm height and 20 mm width created.

Different material properties were defined for chicken breast soft tissue (TABLE I) [15], [16], and for water jet. Equation of State (*EoS*) type $U_s - U_p$ (shock velocity, $U_s$, and particle velocity, $U_p$) has been used for water jet in the finite element model (TABLE II) [17], [18].

TABLE I
CHICKEN BREAST SOFT TISSUE PROPERTIES [15], [16]

| Density (Kg/m³) | Young's Modulus (Pa) | Poisson's Ratio |
|---|---|---|
| 1121 | 25000 | 0.495 |

TABLE II
WATER PROPERTIES [17], [18]

| Density (Kg/m³) | $C_o$ (m/s) | s (Pa.s) | Gamma $_o$ |
|---|---|---|---|
| 983 | 1450 | 1.5 x 10-8 | 0 |

In next step, the two parts were assembled by creating instances from parts; Chicken and Water as dependent, and then, the water has been positioned in the centre of the sample by using translate instance function.

Moreover, a step has been created for the model as Dynamic, Explicit with a time period 0.1 and time scaling factor 1. After that, field output was created for the step selecting the following variables: *S* for stress, *PE*, *PEEQ* and *LE* for strain, *U* for displacement, *V* for velocity, *A* for acceleration, *RF* for reaction forces and moments, and *CSTRESS* for contact stress.

A surface to surface contact (Explicit) has been created as an interaction type between the water jet and the sample with a kinematic contact method as a mechanical constrain formulation.

For the load, a boundary condition was created as pinned that hold the two lower edges of the sample. Next, a predefined field was created to apply a velocity for the water jet toward the sample, which differs for each nozzle diameter. The velocity values were taken from experiment1. After that, the two parts have been meshed independently with approximate global size of 0.082 and the element shapes for both parts were Quad-dominated with a structured technique. Next, the element type for both pieces was "CPE4R", which is a 4-node bilinear plane strain quadrilateral with reduced integration and viscous hourglass control that has a linear geometric order with enabled element deletion.

### III. RESULT AND DISCUSSION

After analysing the results, comparison charts have been generated from experiment 1, experiment 2 and simulation to study the effect of various nozzle diameters on dissecting soft tissue.

#### A. Experimental results

*Experiment 1*
In the first experiment, velocity values were decreasing when the nozzle diameter increased. Thus, resulted in a deeper cut with the same cutting time, 20 seconds, that reaches to total cut (10) from 0.8 mm nozzle diameter and onward (Fig. 4). However, the cutting quality was best for the smallest nozzle diameter. The cutting quality, as shown in Fig. 4, started to reduce from 0.7 mm, which at this point both cutting quality and cutting depth were crossing that concludes the 0.7 mm is a critical point for the chicken breast soft tissue target.

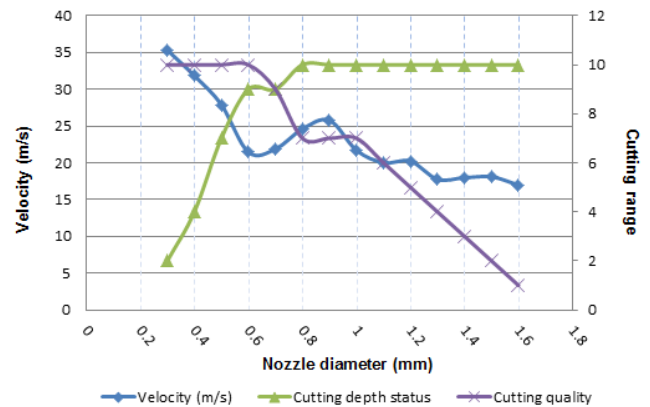

Fig. 4 Comparison chart for velocity, cutting depth and quality for all nozzle diameters

Additionally, 0.6 mm nozzle had better quality and cutting depth compared to the remaining diameters, while the velocity for 0.6 mm and 0.7 mm were almost the same (Fig. 4), which may be caused by turbulent flow of water jet.

Moreover, smaller nozzle diameters; from 0.3 mm to 0.6 mm, had more precise cutting quality compared to bigger nozzle diameters. The bigger nozzles increase the width of the cut which is significantly bigger than the actual diameter of the nozzle due to the scattered water generated from hitting the sample. As shown in Fig.4, the velocity was going down, oppositely, the flow rate is rising, which caused an unpredicted dissecting width to the target sample.

In conclusion, continuous water jet is a good solution to cut soft tissues. However, the quantity of liquid consumed and the time needed to cut the soft tissue might considered as a disadvantage to dissect pituitary gland tumour because of the tumour critical location near the optical nerve and between blood vessels. Therefore, experiment 2 has been performed to study the effect of variable pressure with pulsed liquid jet.

*Experiment 2*
This experiment has been done by using pulsed liquid jet with various: distances, angles and pressures. Hence, to study their effect on peeling soft tissue chicken breast soft tissue.

After preparing the experiment and measuring the peeling depth that has been created from applying the liquid jet. Therefore, results obtained for the samples dissecting depths have been plotted in two graphs providing a clear comparison between each parameter. The results have been summarized for 10º and 30º angles (Fig. 5 and Fig. 6). Each point in Fig. 5 and Fig. 6 graphs represents a distance in millimetres between the nozzle tip and the sample. The pressure set 1.0 bar for the first try (1P) and 2.0 bar for the second try (2P). As a result, each nozzle diameter has eight markers with twenty four points in total that represents the total number of test done in this experiment.

From Fig. 5 and Fig. 6, 0.9 mm nozzle values had almost the highest dissecting depth compared to 0.5 mm and 0.7 mm, which were due to the thickness of the liquid jet that provides more flow and bigger splash compared to 0.7 mm and 0.5 mm nozzles. In addition, the 2 mm distance with 30º angle had the highest value for 0.9 mm nozzle that resulted in a thick cut.

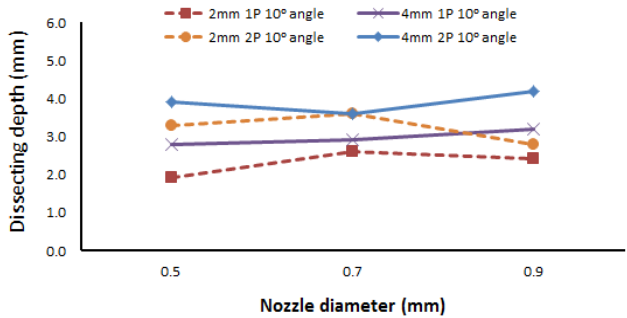

Fig. 5 Dissecting depth values for 10º angle parameter in experiment 2

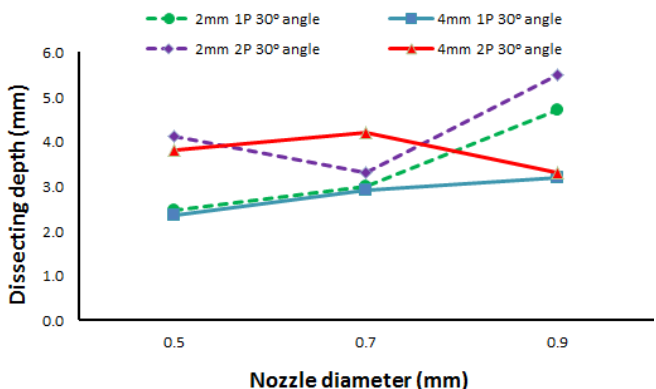

Fig. 6 Dissecting depth values for 30º angle parameter in experiment 2

Furthermore, the sample for 0.7 mm nozzle with 4mm, 2.0 Pa, 10º angle setting creates a thin peeled layer of the tissue that reduced the dissecting depth, which effect the penetration to match closely with the depth of 2mm, 2.0 Pa, 10º angle setting.

Moreover, it has been observed that when the pressure increased the peeling depth get bigger. In addition, 0.7 mm nozzle had a steady increment compared to the other two nozzles. Moreover, the penetration distance was greater for 30º angle because the liquid jet can go deeper into the soft tissue. Therefore, the dissecting width was greater because of the scattered liquid.

Finally, it is evident from experiment 2 that the soft tissue property can affect the peeling depth. With this in mind, the pituitary gland tumour has varying size and hardness that required variable pressure setting, which can be adjusted during the surgery. Even more, the simplicity and the size of the used device in experiment 2 provides an advantage compare to Ogawa et al. [10] device, which is bigger in size and uses an expensive laser beam device to generate the pulsed liquid jet.

*B. Numerical results*

After creating 2D models and 3D models, the analyses for these models were carried out, and then the results were detailed afterward. Thus, a comparison charts were made from these results to discuss the effect of various nozzle diameter on cutting a soft tissue of chicken breast.

The 2D and 3D models focus on the time required for each nozzle to penetrate the sample, the performance of water velocity when cutting the sample, and the dissecting width created from the water jet. Additionally, a plot has been generated for 2D models to show the distance that water splash reaches after hitting the sample.

*2D models*

To study the influence of nozzle diameters on cutting penetration, a plot as shown in Fig. 7, has been created. The plots illustrate the simulation time to cut through a 1.0 mm chicken breast soft tissue. It shows that 0.1 mm of thickness was dissected at the same time for all nozzles. After that, the lines were diverse except for 0.3 mm and 0.4 mm nozzles which dissected the same depth until near 0.8 mm of thickness, where they separated. Similarly, 0.6 mm and 0.7 mm nozzles cut the sample at the same time till 0.4 mm of displacement, where they separated and continue to penetrate the sample in parallel line with the shortest time, unexpectedly, for 0.7 mm nozzle. On the other hand, 0.5 mm nozzle had its own cutting time that started from the sample thickness of 0.1 mm, and then increased the dissecting depth gradually until fully penetrated the sample.

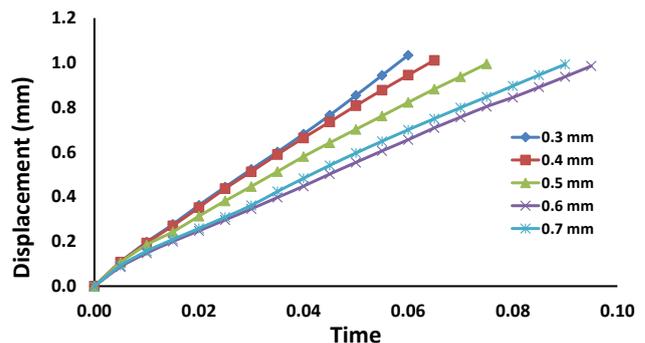

Fig. 7 Penetration depths related to each nozzle diameter

However, the water jet velocity for each nozzle (Fig. 8) had dropped to more than half its value at the same time, 0.01s, which is the time to dissect 0.1 mm. Moreover, the three smallest nozzle diameters had rose their velocity 2 m/s after dissecting 0.1 mm, which means that 0.5 mm nozzle is considered as a critical diameter for the water jet. Furthermore, the two 0.6 and 0.7 mm nozzle diameters continued to drops more of their velocity, then remained approximately at the same points until penetrating the sample.

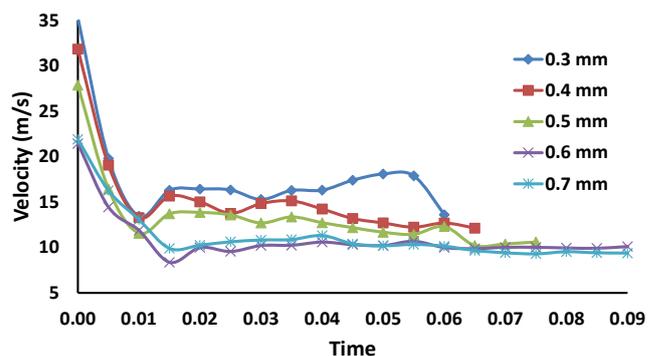

Fig. 8 Velocity effect on each nozzle diameter when dissecting the sample

Next, a graph has been created by merging the data gained from the sample path line for each nozzle diameter to demonstrate the penetration width formed by each nozzle diameter (Fig. 9).

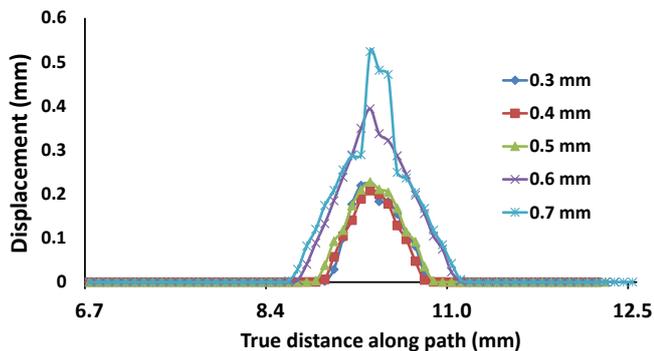

Fig. 9 An overview chart for penetration width created by each nozzle diameter

To make a clear measurement, a new plot has been generated (Fig. 10) that shows the X-axis from 8 mm to 12 mm long, which can be extracted that the maximum width of 2.5 mm was created by 0.7 mm nozzle followed by 0.6 mm nozzle that penetrated 2.2 mm of width. Meanwhile, the three smallest nozzle diameters had almost the same cutting width of 1.9 mm; therefore, 0.5 mm nozzle can be consider a better choice since it is easier to create and has the same effect on penetration width.

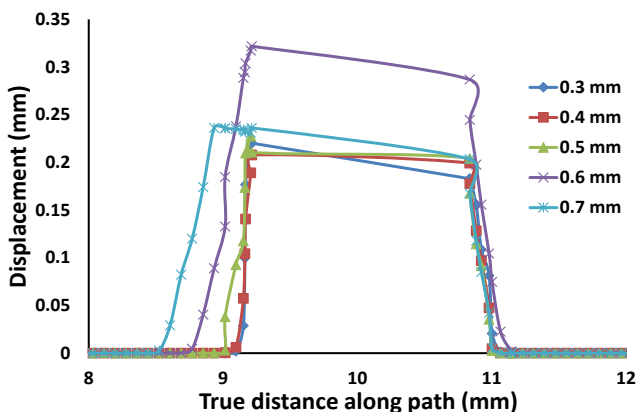

Fig. 10 A close look chart for penetration width created by each nozzle diameter

Moreover, a water splash chart has been generated that shows the maximum distance the water particle had travelled after hitting the sample for each nozzle diameter (Fig. 11). As a result, it has been observed that the distance of the water splash is not linked to the nozzle diameter, except for the biggest diameter that had the extreme water splash. Hence, 0.7 mm nozzle had the greatest total splash distance that equal to nearly double its diameter from the right side (1.36 mm) and the same for the left side (1.32 mm). Surprisingly, 0.4 mm nozzle diameter had the next wider splash that spreads 1.02 mm from the right side and 1.33 mm from the left side. Afterward, the splash reduced in size for 0.6 mm nozzle, followed by 0.5 mm and 0.3 mm nozzle diameters that had approximately the same splash distance. Furthermore, only 0.5 mm nozzle and 0.7 mm nozzle had an even splash distance that were equally travelled in both directions.

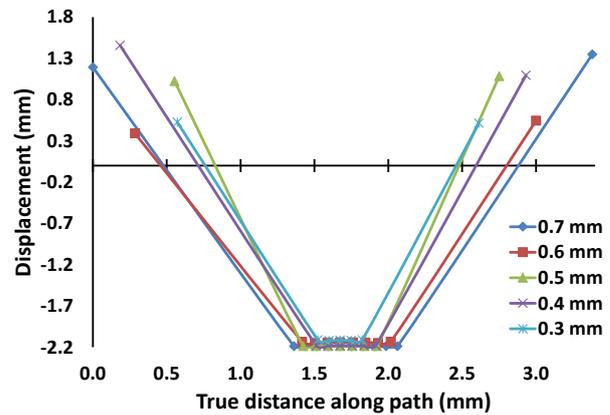

Fig. 11 Water splash traveling distance for each nozzle diameter

*C. Summary*

The dissecting width from both experiments and simulation are affected by the nozzle diameter, as larger nozzle will get a wider cut but takes more time to reach the desired point. In addition, the liquid consumed volume will be higher for bigger nozzle diameter that must be taken into account when working in a small and narrowed area; for instance, pituitary gland area, where it needs smaller liquid jet diameter for more precise tumour dissecting in a quick time and less liquid volume.

After analysing the results from both experiments and simulation, 0.5 mm nozzle is the best diameter to dissect pituitary gland tumour because it has steady depth with an even water splash that creates constant cutting width for less water volume.

Moreover, the size of the device and easily controlling of the liquid pressure are the advantages of using water pump compared to using laser to create the liquid jet that had been used in Ogawa et al. [1] study, which is bigger in size and needs to control the laser power to vary the liquid pressure. Furthermore, the device in experiment 2 can be modified to produce lower pressure with different pulsed ratio, which can be done for future work.

In addition, there are several benefits of using pulsed liquid jet compared to current surgical methods of dissecting pituitary gland tumour. Accordingly, preserving fine blood vessels and optical nerve located near the pituitary gland, less harm to surrounding anatomical structures, and provides a safe and easy use to the surgeon that eliminate damages to pituitary gland and neighbouring tissues.

IV. CONCLUSION

Since the current methods to dissect pituitary gland tumour have many disadvantages, designing a new medical device to dissect the tumour in less harm to the patient with minimal invasive procedure that provides a high success rate is essential to reduce the complexity of cutting the lesion that lead to physician faults, which can cause more medical complications to the patient. Therefore, liquid jet has the ability to remove the tumour with no side effects when designed with a safeguard configuration that avoid any damage to blood vessels and optical nerve by limiting the jet pressure. In addition, it provides a precise cutting depth with various nozzle angles that chosen according to the surgeon requests.

Moreover, results from the continuous water jet experiment has demonstrates that the cutting quality and depth were influenced by the nozzle diameter when fixing the water inlet pressure, in which, the cutting quality were decreased when the jet diameter get bigger. In contrast, the cutting depth increased proportional to the nozzle diameter until 0.8 mm nozzle, where the water jet penetrated the sample. Next, pulsed liquid jet was conducted with more parameters that improve the dissecting depth with less liquid volume compared to continuous water jet experiment. Furthermore, the simulation shows clear outcomes to the effect of various nozzle diameters related to the dissecting width and the penetration time that is close to the experiment results, which gives estimated values when selecting the jet diameter.

Due to the project's time limitation, the number of nozzle diameter variables for experiment 2 has been reduced to three diameters. Also, additional work is required for pulsed liquid jet experiment to control the jet pulse rate, provide more variables for the jet pressure, and test nozzle diameter with the same length required during the surgery. In addition, simulate a finite element model with Eulerian type for water jet to provide outcomes that are more precise.

Finally, pulsed liquid jet technology will replace current surgical methods in the future, not only for hypophysis tumour, but also for other body parts because it consume less water, which improves protecting small body structures with better accessibility that affords stable well cutting ability in narrowed area [19].